\documentclass[11pt,preprint]{aastex}

\usepackage{natbib}

\newcommand{\kms}{\,km~s$^{-1}$}

\newcommand{\Msun}{\mbox{\,$M_{\odot}$}}

\def\spose#1{\hbox to 0pt{#1\hss}}
\def\simlt{\mathrel{\spose{\lower 3pt\hbox{$\mathchar"218$}}
     \raise 2.0pt\hbox{$\mathchar"13C$}}}
\def\simgt{\mathrel{\spose{\lower 3pt\hbox{$\mathchar"218$}}
     \raise 2.0pt\hbox{$\mathchar"13E$}}}

\slugcomment{Accepted to AJ}

\shorttitle{A Counter-Rotating Core in NGC 770}
\shortauthors{Geha et al.}

\begin{document}


\title{NGC 770: A Counter-Rotating Core in a Low-Luminosity Elliptical Galaxy}


\author{M.\ Geha\altaffilmark{1}}
\affil{The Observatories of the Carnegie Institute of Washington,
        813 Santa Barbara Street, Pasadena, CA~91101}
\altaffiltext{1}{Hubble Fellow}
\email{mgeha@ociw.edu}

\author{P.\ Guhathakurta}
\affil{UCO/Lick Observatory, University of California,
    Santa Cruz, 1156 High Street, Santa Cruz, CA~95064}
\email{raja@ucolick.org}

\and

\author{R.\ P.\ van der Marel}
\affil{Space Telescope Science Institute, 3700 San Martin Drive, Baltimore,
MD~21218}
\email{marel@stsci.edu}


\begin{abstract}
\renewcommand{\thefootnote}{\fnsymbol{footnote}}

We present evidence for a counter-rotating core in the low-luminosity
($M_B = -18.2$) elliptical galaxy NGC~770 based on internal stellar
kinematic data.  This counter-rotating core is unusual as NGC~770 is
not the primary galaxy in the region and it lies in an environment
with evidence of on-going tidal interactions.  We discovered the
counter-rotating core via single-slit Keck\footnote{Data presented
herein were obtained at the W.\ M.\ Keck Observatory, which is
operated as a scientific partnership among the California Institute of
Technology, the University of California and the National Aeronautics
and Space Administration.  The Observatory was made possible by the
generous financial support of the W.\ M.\ Keck Foundation.}/ESI
echelle spectroscopy; subsequent integral field spectroscopy was
obtained with the Gemini/GMOS IFU.  The counter-rotating region has a
peak rotation velocity of 21\kms\ as compared to the main galaxy's
rotation speed of greater than 45\kms\ in the opposite direction.  The
counter-rotating region extends to a radius of $\sim4''$ (0.6\,kpc),
slightly smaller than the half-light radius of the galaxy which is
$5.3''$ (0.8\,kpc).  The photometry and two-dimensional kinematics
suggest that the counter-rotating component is confined to a disk
whose scale height is less than $0.8''$ (0.1\,kpc).  The major-axis of
counter-rotation is misaligned with that of the outer galaxy isophotes
by $15^{\circ}$.  We compute an age and metallicity of the inner
counter-rotating region of $3 \pm 0.5$\,Gyr and [Fe/H] = $0.2\pm
0.2$\,dex, based on Lick absorption-line indices.  The lack of other
large galaxies in this region limits possible scenarios for the
formation of the counter-rotating core. We discuss several scenarios
and favor one in which NGC~770 accreted a small gas-rich dwarf galaxy
during a very minor merging event.  If this scenario is correct, it
represents one of the few known examples of merging between two
dwarf-sized galaxies.

\end{abstract}




\section{Introduction}\label{intro_sec}

Kinematically-distinct cores (KDCs) occur in $\sim10\%$ of elliptical
galaxies; a few tens of such systems are currently known
\citep{sur95,meh98,wer02, dez02, der04,ems04}.  The
kinematically-distinct regions have sizes ranging in radius from 0.06
to 2\,kpc and have been observed in galaxies which reside in
environments ranging from loose groups to dense galaxy clusters.
Counter-rotating cores are a subset of KDCs and provide the strongest
evidence that these regions are formed during merging events.
\citet{bal90} investigated the formation of counter-rotating cores via
the dissipationless merging of unequal mass elliptical galaxies.  In
this scenario, the dense core of the smaller galaxy survives the
merger and sinks to the center of the primary galaxy via dynamical
friction.  For certain merging geometries, the surviving core will be
kinematically misaligned and will occasionally counter-rotate with
respect to its host galaxy.  Merging/interactions are not necessarily
required to explain kinematically-distinct cores in all cases.
\citet{sta91} demonstrated that projection effects in a
dynamically-smooth triaxial system were able to explain a KDC in
NGC~5982 which rotates about its minor-axis.  Projection effects,
however, do not easily explain cores with counter-rotation about the
major axis.  

The theories described above cannot naturally explain the presence of
faint stellar disks detected on the same scale as the KDCs in several
galaxies.  In these cases, the more likely formation scenario is the
merging or interaction of two separate systems with available gas
reservoirs.  During the interaction, a disk is formed via the
dissipative infall of gas, which eventually cools to form stars
\citep{her91}.  \citet{car97} presented photometric evidence for
underlying disks in roughly half of the 15 elliptical galaxies with
kinematically-distinct cores they imaged with {\it Hubble Space
Telescope\/} Wide Field Planetary Camera~2.  These authors find no
significant color difference between the disk and main galaxy body,
although evidence of color differences and stellar population
gradients has been presented in other systems.  \citet{meh98} and
\citet{sur95} find evidence for line-strength gradients in three
separate systems in the sense that the kinematically-distinct cores
have enhanced $\alpha$-element to iron abundance ratios relative to
the main galaxy, but similarly old ages. They interpret these
gradients as evidence of rapid star formation during the formation of
the KDC, however, similar gradients have been observed in ellipticals
galaxies which do not host KDCs \citep{fis96,meh03}.  In addition to
the presence of disks, strong line-strength gradients argue further in
favor of dissipational merging since dissipationless processes tend to
weaken such gradients.

The majority of known KDCs reside in elliptical galaxies brighter than
$M_B < -19$.  \citet{der04} recently presented evidence for KDCs in
two dwarf elliptical galaxies ($M_V \sim -17$) found in galaxy-rich
groups.  The cores of these two dwarf galaxies rotate in the same
direction as the main galaxy body, but with a slower rotation speed.
These authors propose an alternative formation mechanism to the
merging scenario in which KDCs are formed via flyby interactions
between the dwarf and a more massive galaxy.  Angular momentum is
transfered to the outer envelope of the dwarf during these
interactions which results in a kinematic decoupling between the outer
envelope and inner core.  Such a scenario is favored in dense galaxy
environments where the probability of dwarf-dwarf galaxy merging is low
due to the large relative velocities between galaxies.

\begin{figure}
\vskip -18cm
\plotone{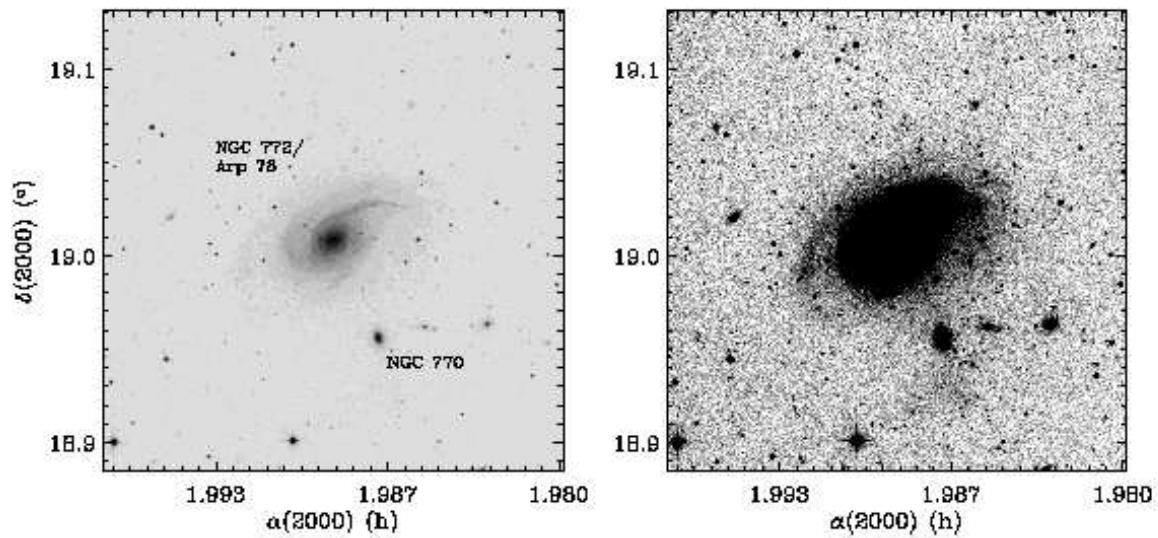}
\caption{The environment of the low-luminosity elliptical galaxy
NGC~770.  ({\it Left\/})~Digitized Palomar Optical Sky Survey-II image
centered on the parent spiral galaxy NGC~772/Arp~78; the satellite
NGC~770 is located $4'$ to the SW of the primary spiral.  The image is
$15'\times 15'$.  ({\it Right\/})~Same image displayed at higher
contrast to highlight the faint tidal features extending from both
NGC~772 and NGC~770.
\label{fig_dss}} 
\end{figure}

NGC~770 is a low-luminosity elliptical ($M_B = -18.2$) companion to
the large spiral galaxy NGC~772, and is the brightest satellite galaxy
identified in this system by \citet{zar97}.  The parent spiral galaxy
NGC~772 has an absolute magnitude $M_B = -21.6$ and is listed in the
Atlas of Peculiar Galaxies \citep[Arp 78;][]{arp66}.  This system has
been included in many studies of interacting systems
\citep[e.g.,][]{pig01,tut97,elm91,lau89} due to a prominent asymmetric
spiral arm and faint trails of surrounding material as seen in
Figure~\ref{fig_dss}.  

Here, we present evidence for a counter-rotating core in the satellite
galaxy NGC~770.  The counter-rotating core in NGC~770 is particularly
interesting as it is one of the faintest elliptical galaxies known to
host a counter-rotating core and resides in an environment which shows
evidence of recent interactions.  This system may provide unique clues
to the formation of counter-rotating cores and the presumed role of
dissipation in the merging process.

This paper is organized as follows.  Observations and reduction
procedures for Keck/ESI and Gemini/GMOS IFU data are discussed in
\S\,\ref{n770_data}.  We present surface photometry and argue for the
presence of a central disk in \S\,\ref{sec_disk}.  In
\S\,\ref{sec_vp}, we present the one- and two-dimensional kinematic
profiles/maps for NGC~770.  We measure the stellar
absorption-line-strengths in \S\,\ref{sec_lick} and determine the age
and metallicity of the counter-rotating region.  In
\S\,\ref{sec_environ}, we discuss the environment of NGC~770 and
evidence for recent and on-going interactions with the primary galaxy.
Finally, in \S\,\ref{sec_discuss}, we discuss possible scenarios for
the formation of the counter-rotating core in NGC~770.

A true distance modulus to NGC~770 of $(m - M)_0 = 32.6$, i.e.,~a
distance of 32.9~Mpc, is adopted throughout this paper.  The
distance is determined from the systemic velocity of the primary
galaxy in this system, NGC~772, of 2468\kms\ \citep{zar97} assuming
$H_0=75$\kms\,Mpc$^{-1}$.  A line-of-sight extinction value of $A_V =
0.24$ is adopted from \citet*{sch98} assuming a standard Galactic
reddening law with $R_V = 3.1$.

\section{Observations and Data Reductions}\label{n770_data}

NGC~770 was first confirmed as a satellite galaxy to NGC~772 by
\citet{zar97} via spectroscopy.  The satellite galaxy NGC~770 was
targeted for high-resolution spectroscopy as part of our survey of
dwarf and low-luminosity elliptical galaxies in a variety of environments
\citep*{geh03}.  As seen in Figure~\ref{fig_dss}, NGC~770 lies at a
projected distance of only $3.5'$ (33\,kpc) from its parent galaxy.  The
basic properties of NGC~770 are listed in Table~1.

\subsection{Single-Slit Keck/ESI Echelle Spectroscopy}\label{subsec_esired}

\begin{figure}
\plotone{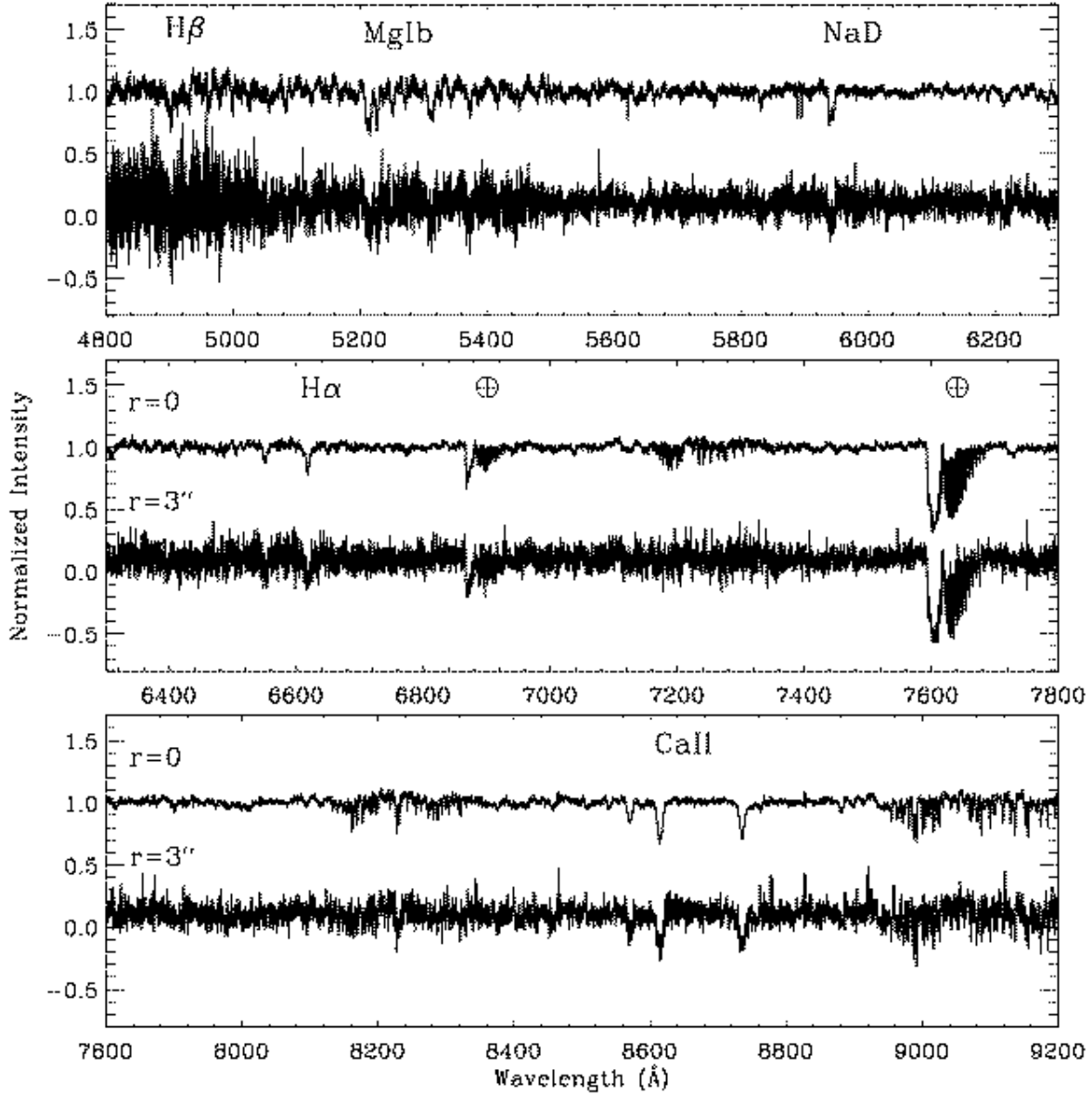}
\caption{Keck/ESI echelle spectra of NGC~770 covering the continuous
wavelength region $4800$--$9200\mbox{\AA}$.  Spectra
are shown for the galaxy center ($r=0''$) and the
edge of the counter-rotating region ($r=3''$).  The spectra are
binned spatially to $0.9''$, the seeing FWHM at the time
of the observations, but have not been smoothed in the spectral direction.
A few of the important stellar absorption features are indicated,
along with the atmospheric A and B absorption bands.
\label{fig_spec}} 
\end{figure}

Spectroscopic observations of NGC~770 were obtained on 2001 October 12
using the Keck~II 10-m telescope and the Echelle Spectrograph and
Imager \citep[ESI;][]{she02}.  The instrument was used in the
echellette mode with continuous wavelength coverage over the range
$3900$--$11000\mbox{\AA}$ across 10~echelle orders
with a spectral resolution of $R \equiv \lambda / \Delta\lambda \sim
10,000$; our analysis is restricted to the spectral range
$4800$--$9200\mbox{\AA}$ over which useful information
can be extracted.  The spectra were obtained through a $0.75'' \times
20''$ slit, resulting in an instrumental resolution of 23\kms\
(Gaussian sigma) over the entire spectrum.  NGC~770 was observed for
$3\times 600$\,s under seeing conditions of $0.9''$.  The slit was
positioned on the major axis of the galaxy (as determined from Digital
Sky Survey images) at a position angle of $+5^{\circ}$ on the sky.
The ESI data were reduced using a combination of IRAF echelle and
long-slit spectral reduction tasks
\citep*{geh02}.  The final combined spectrum was rebinned into
logarithmic bins with 11.4\kms\ per pixel for the kinematic analysis
(\S\,\ref{subsec_1d}), and linear bins with $0.2 \mbox{\AA}$ per pixel
for the line-strength analysis (\S\,\ref{sec_lick}).  In
Figure~\ref{fig_spec}, we compare the full spectrum extracted in a
$0.9''$ aperture at the center of the galaxy to a similarly sized
aperture $3''$ off center.

Due to the short ESI slit ($20''$) and the spatial extent of NGC~770,
the sky spectrum determined at the extreme edge of the slit contains
some galaxy light.  We estimate the level of this contamination at
$\sim 2\%$ in the galaxy center, increasing to $\sim20\%$ at a radius
of $r=4''$.  We have tested that galaxy contamination does not affect
our results by repeating the analysis using non-local sky subtraction
created by combining the sky spectra from unrelated galaxies observed
through the same ESI set-up.  The resulting kinematic profiles and
line-strength values were within the 1-sigma error bars.  The results
presented in this paper are based on local sky subtraction as this
provided the smallest sky-subtraction residuals near bright night-sky
lines.

The mean line-of-sight velocity ($v$), velocity dispersion ($\sigma$),
and higher-order Gauss-Hermite moments ($h_3$ and $h_4$) were
determined as a function of radius using a pixel-fitting method
described in \citet{vdm94}.  The velocity profile was recovered using
template stars ranging in spectral type from G8 III to M0 III
convolved with an appropriate kernel.  The best-fitting template, the
K0III star HD~6203 ($\rm[Fe/H] = -0.3\,$dex), was used to recover the
profiles presented here.  This stellar template, convolved with the
best-fitting Gaussian kernel, is fit to the observed spectra to
determine $v$ and $\sigma$; these parameters are than held fixed while
the higher-order deviations from Gaussian are determined.  The galaxy
data were spatially binned to ensure a minimum signal-to-noise level
of $\rm S/N = 10$ per bin at all radii and a minimum bin-size greater
than or equal to that of the seeing FWHM during the observations
($0.9''$).  The signal-to-noise ratio in the spatial bins inside
$r=3''$ is $\rm S/N \ge 25$.  As demonstrated in \citet{geh02},
internal velocity dispersions from this observing setup can be
measured down to the instrumental resolution of 23~km~s$^{-1}$~with an
accuracy of 1\% at our minimum S/N level.  The systemic radial
velocity of NGC~770 was determined from the central velocity
measurement and subtracted from the velocity profile.  The corrected
heliocentric systemic velocity is $v_{\rm sys} = 2538 \pm 5$\kms, in
agreement with previous measurements by \citet{zar97} $v_{\rm sys} =
2543 \pm 22$\kms. The one-dimensional Keck/ESI single-slit kinematic
profiles are shown in Figure~\ref{fig_vpesi}.

\subsection{Gemini/GMOS IFU Spectroscopy}\label{subsec_gmosred}

Follow-up two-dimensional spectroscopy of NGC~770 was carried out with the
Gemini Multiobject Spectrograph \citep[GMOS;][]{hoo04} Integral Field
Unit \citep[IFU;][]{mur03} at the Gemini North Telescope on 2003
October 29.  The IFU data were obtained in two-slit mode providing a
$5''\times 7''$ field-of-view consisting of 500 $0.2''$ diameter
hexagonal fibers; an additional 250 fibers are dedicated to sky $1'$
from the science field.  We placed the long axis of the IFU along the
major axis of the galaxy at a position angle of $+5^{\circ}$.  We used
the G400 grating and CaT filter which provide wavelength coverage in
the region $7800$--$9200\mbox{\AA}$ with a dispersion of
0.67$\mbox{\AA}$ per pixel.  The spectral resolution is estimated to
be 2.8$\mbox{\AA}$ FWHM, based on the $0.2''$ fiber diameters, a
pixel scale of $0.072''$ and an anamorphic magnification factor of
0.645.  NGC~770 was observed for $2\times 3300$\,s.  The seeing FWHM
was $0.7''$, estimated from acquisition images taken just prior to the
science observations.  The data were reduced using the available
Gemini IRAF package\footnote{Available at
http://gemini.edu/sciops/instruments/gmos/gmosIndex.html}.  The final
three-dimensional data cube has $0.1''$ spatial pixels and 25\kms\
spectral pixels.

The mean line-of-sight velocity and velocity dispersion were
determined for the GMOS IFU data in the Ca II triplet region
($8100-8800\mbox{\AA}$) using the same pixel-fitting
method described above \citep[\S\,\ref{subsec_esired};][]{vdm94}.  We
determine the optical center of the galaxy from the reconstructed IFU
image and set this point equal to ($x$,~$y$)~=~(0,~0)
in the kinematic maps.  To
achieve a minimum signal-to-noise level of $\rm S/N = 10$ per
two-dimensional bin, we spatially rebinned the data using the Voronoi
2D binning algorithm of \citet{cap03}.  A template star taken through
the same GMOS observing set-up was not available.  Instead, velocities
were recovered using the same ESI stellar template described in
\S\,\ref{subsec_esired}.  The ESI template was rebinned to match the
spectral resolution of the GMOS IFU.  The measured velocity
dispersions are then the quadrature sum of the internal galaxy
velocity dispersion and the GMOS instrumental profile.  We determine
the GMOS instrumental profile by setting the central velocity
dispersion equal to that measured in the ESI observations.  The
instrumental profile is subtracted in quadrature from the measured
velocity dispersions to give the internal velocity dispersion of the
galaxy.  Because it is difficult to determine the shape of the
instrumental profile exactly (here we assume it is Gaussian), we do
not attempt to recover the higher-order Gauss-Hermite moments for the
GMOS IFU data.  The resulting mean velocity and velocity dispersion
maps for NGC~770 are shown in Figures~\ref{fig_gmos_vel} and
\ref{fig_gmos_sig}, respectively.

\subsection{Keck/ESI Imaging}\label{subsec_imgred}

Keck/ESI $V$-band images of NGC~770 were obtained during the same run
as the single-slit spectral observations described in
\S\,\ref{subsec_esired}.  In imaging mode, ESI has $0.154''$ pixels
and a $2'\times 8'$ field-of-view.  The seeing FWHM was $0.9''$ in our
$2 \times 60$\,s $V$-band exposures.  The images were bias subtracted,
flat fielded using twilight sky exposures, and combined.  The
photometric zeropoint was determined based on standard star fields
taken on a different night from our observations.  The resulting image
is shown in Figure~\ref{fig_sb2d} (left panel).  The surface
brightness profile was determined using the IRAF ELLIPSE isophotal
fitting routine \citep{jed87}.  In Figure~\ref{fig_sb}, radial
profiles are shown for surface brightness ($\mu_V$), ellipticity
($\epsilon$), position angle (PA) and higher-order deviations from a
perfect ellipse ($B_4$).  We fit a Sersic profile to the $V$-band
surface brightness profile of the form $I^{\rm gal}(r) = I_0^{\rm gal}
{\rm exp}[(r/r_0)^{1/n}]$.  The best-fit Sersic profile is determined
by non-linear least-squares fitting to the region $r=1''-25''$; the
best fitting Sersic index for NGC~770 is $n=2.3$.  We have also fit a
seeing-convolved two-dimensional Sersic profile using the GALFIT
fitting algorithm \citep{pen02} in the region $r < 30''$ and recover
the same Sersic index, $n=2.3$.  The half-light effective radius,
Sersic index, and other photometric quantities agree with previous
photometry by \citet*{gut02} and are listed in Table~1.

\section{Photometric Analysis}\label{sec_disk}

\begin{figure}
\epsscale{0.9}
\plotone{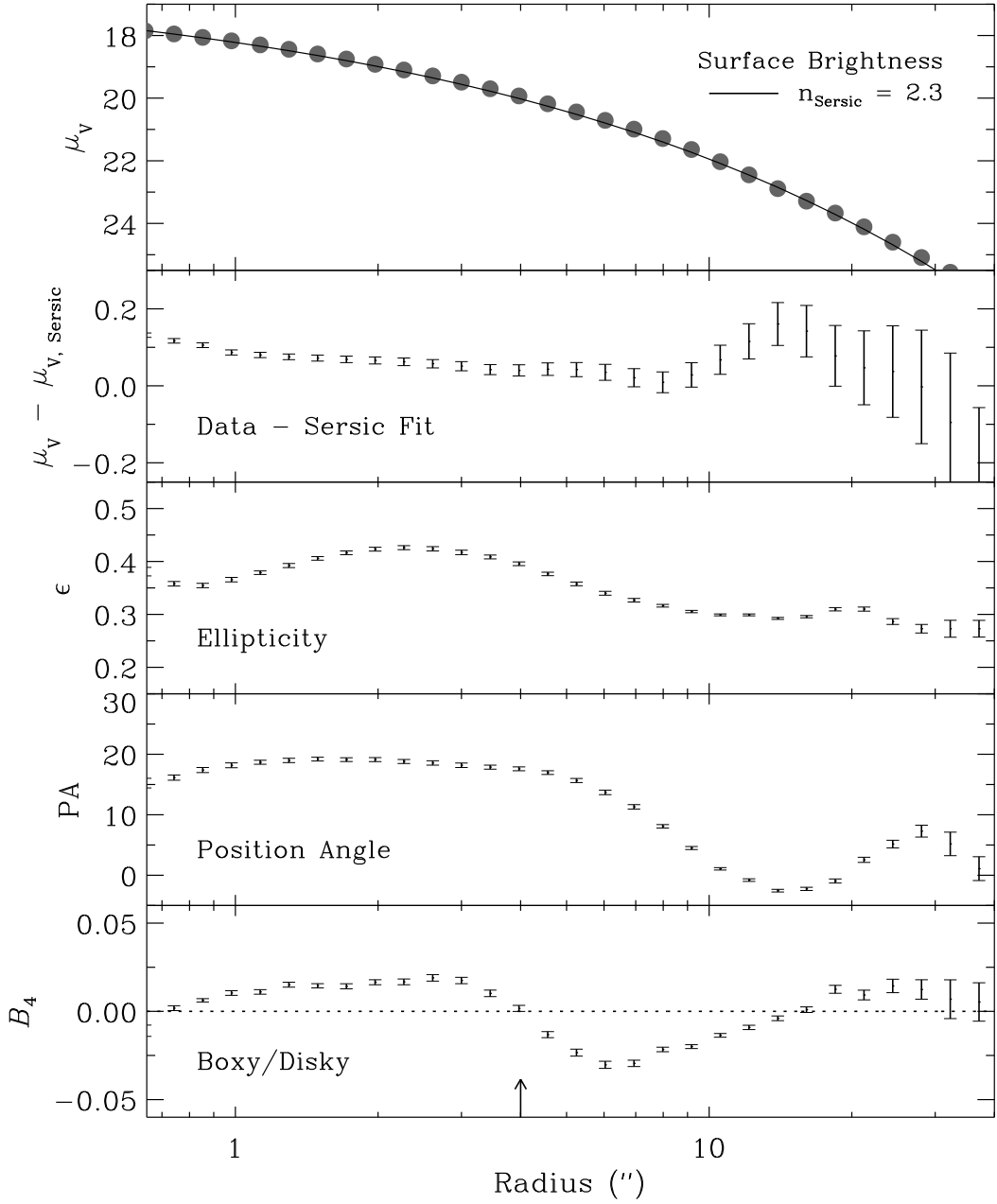}
\caption{Radial profiles for $V$-band surface brightness ($\mu_V$),
residual surface brightness residuals after subtracting the Sersic fit
($\mu_V - \mu_{V, fit}$), ellipticity ($\epsilon$), position angle
(PA) and the boxiness/diskyness parameter ($B_4$) measured from
Keck/ESI imaging ({\it top to bottom\/}).  Positive $B_4$ values
indicate disky photometric contours, while negative values indicate
boxy contours.  In the top panel, the best fitting Sersic profile
($n=2.3$) is plotted over the data.  The radial extent of the
counter-rotating region as measured in the GMOS velocity map is shown
by the arrow at $4''$ in the bottom panel.
\label{fig_sb}} 
\end{figure}

NGC~770 is classified by \citet*{dev76} as an ``E3:'' galaxy, the
colon indicating uncertainty in the classification.  The source of
this uncertainty is due to the shape of the surface brightness profile
of this galaxy, shown in Figure~\ref{fig_sb}.  The profile is not well
fit by either the de~Vaucouleurs $r^{1/4}$ law, nor an exponential
profile.  Rather, the best fitting Sersic index for NGC~770 is
$n=2.3$, intermediate between a $n=1$ exponential profile typical for
a dwarf elliptical and a $n=4$ de~Vaucouleurs law typical of giant
ellipticals.  NGC~770 is half a magnitude brighter ($M_{V,0} = -18.9$)
than the brightest classified dwarf elliptical galaxy in the Virgo
Cluster and three magnitudes brighter than elliptical satellites
NGC~205 and M\,32 in the Local Group.  It is best classified as a
low-luminosity elliptical galaxy.

\begin{figure}
\epsscale{1.}
\plotone{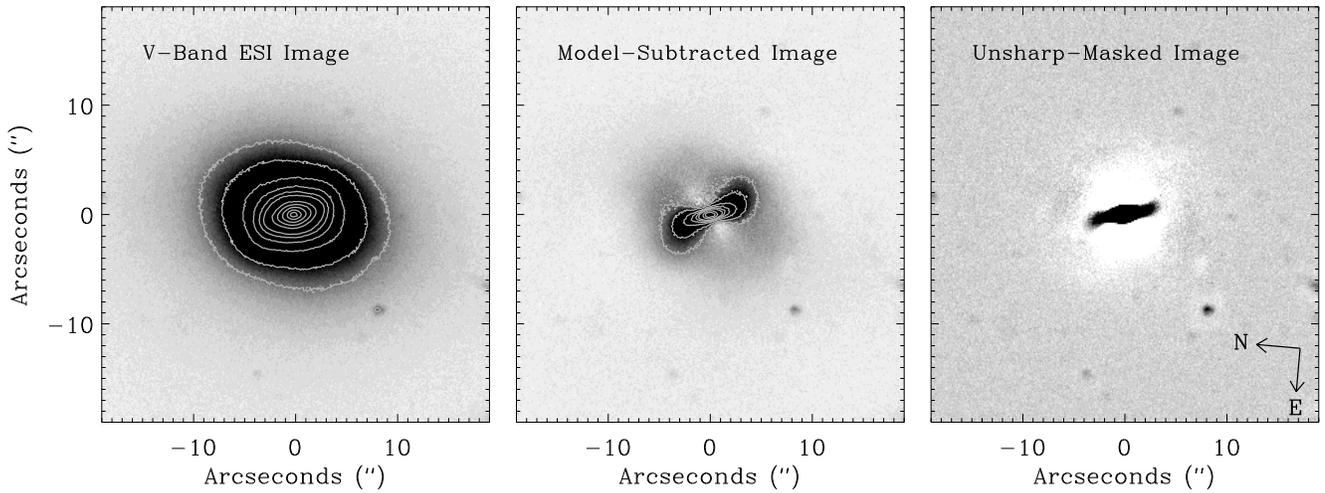}
\vskip 1cm
\caption{({\it Left\/})~Keck/ESI $V$-band image of NGC~770 overlaid
with isophotes in the central region. ({\it Middle\/})~Residual
image created by subtracting a two-dimensional Sersic $n=2.3$ profile
at a constant position angle and ellipticity from the original image.
({\it Right\/})~Unsharp-masked image created by subtracting a smoothed
version of the original image from itself (with a $\rm25\,pixel = 4''$
smoothing length).  The contours in the left and middle panels
represent the same isophote levels.  Each panel covers a $40'' \times 40''$
region centered on the galaxy and is oriented as indicated in the
bottom right corner of the right panel.
\label{fig_sb2d}} 
\end{figure}

Close inspection of the inner photometric contours in
Figure~\ref{fig_sb2d} (left panel), shows elongated, disky central
isophotes and boxy isophotes at larger radii.  This is quantitatively
confirmed by the $B_4$ profile in Figure~\ref{fig_sb} which is
positive (disky) in the central $3''$, and negative (boxy) between
radii $5''$ to $10''$.  To study the substructure implied by the
one-dimensional profile, we plot in Figure~\ref{fig_sb2d} (middle
panel) the result of subtracting a two-dimensional surface brightness
model from the original image.  The two-dimensional model is a Sersic
$n=2.3$ profile at a constant position angle and ellipticity
determined at radii $r\ge4''$.  In the right panel of
Figure~\ref{fig_sb2d}, we present an unsharp-masked image created by
subtracting a boxcar-smoothed image (with a 25\,pixel $= 4''$
smoothing length) from the original.  Both the model-subtracted and
unsharp-masked images suggest a disk in the central region of NGC~770
whose major axis is slightly misaligned with the major axis of the
galaxy at larger radii.  The central disk extends out to a radius of
$3''$--$4''$ (0.5--0.6\,kpc), slightly less than the half-light radius
of the galaxy ($r_{\rm eff} = 5.3''$).  While the vertical extent of
this disk is unresolved in the images ($0.9''$ seeing FWHM), we can
put an upper limit on this quantity of $1''$.  The major axis of the
disky region seen in the model-subtracted and unsharp-masked image is
misaligned by $15^{\circ}$ with respect to the major axis of the main
galaxy body, consistent with the isophotal twist between the inner and
outer regions seen in the position angle profile in
Figure~\ref{fig_sb}.  To estimate the fraction of light contained in
the disky region, we compare the flux in the model-subtracted image to
that of the original.  The model was scaled such that no negative flux
was present in the inner $10''$ of the model-subtracted image.  The
model-subtracted image contains 6\% of the total galaxy light;
isolating the elongated disk region inside a radius of $4''$
represents 3.5\% of the total galaxy light.  We show below that the
inner disky region is co-spatial and aligned with the counter-rotating
core.

\section{Stellar Kinematics}\label{sec_vp}

\subsection{One-Dimensional Kinematics}\label{subsec_1d}

\begin{figure}
\epsscale{0.9}
\plotone{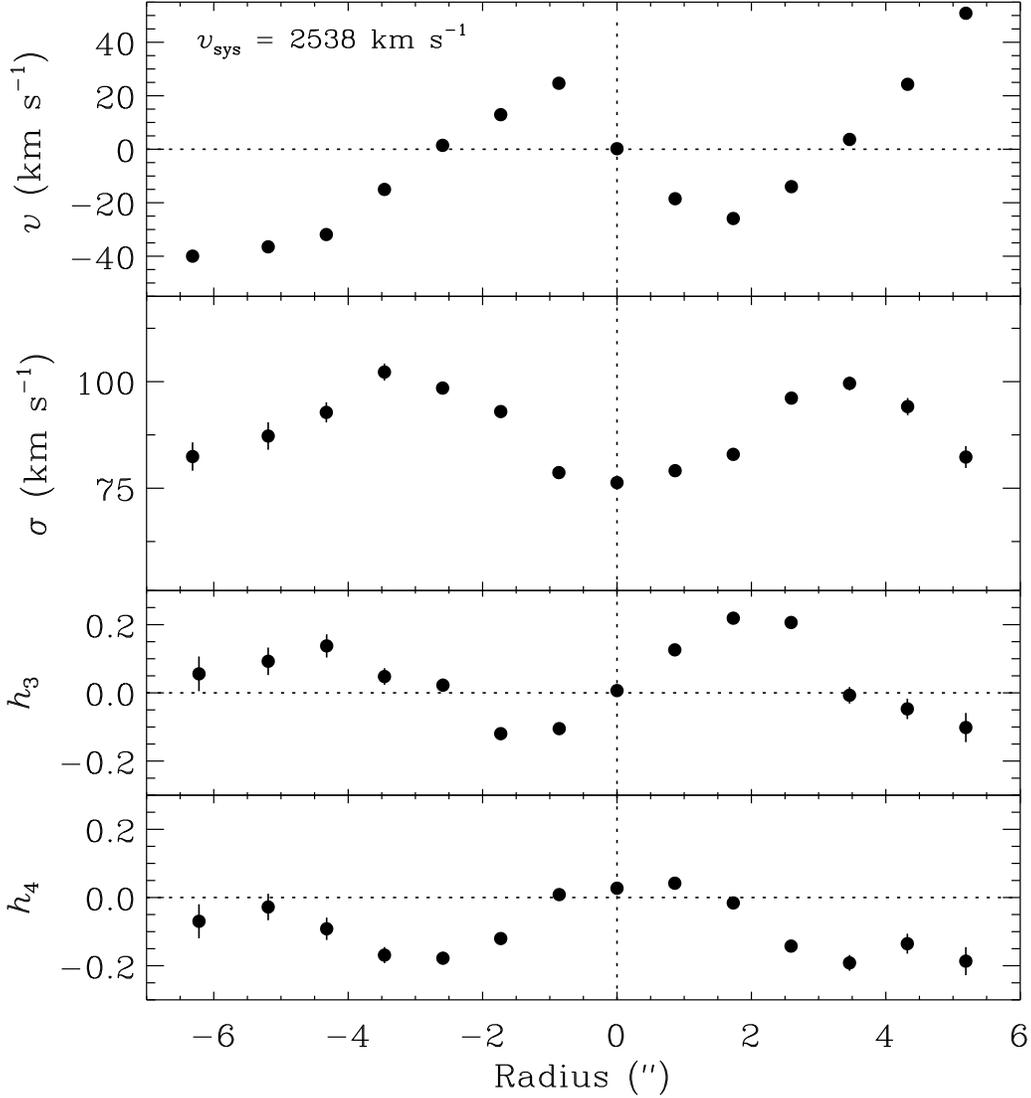}
\vskip -1.5cm
\caption{({\it Top to bottom\/})~Radial profiles of the mean
line-of-sight velocity $v$, velocity dispersion $\sigma$, and
higher-order Gauss-Hermite moments, $h_3$ and $h_4$, measured via
Keck/ESI single-slit observations.  The slit was placed along the
photometric major axis of the main body of NGC~770, which is misaligned
with respect to the kinematic major axis of the
counter-rotating region by $15^{\circ}$.  The velocity
profile reveals a kinematically-distinct core rotating in a direction
opposite to that of the main galaxy body.  At the distance of NGC~770,
$1''=160$\,pc.  One-sigma error bars are plotted in all panels, but
are often smaller than the plotted data points.
\label{fig_vpesi}} 
\end{figure}

The counter-rotating core in NGC~770 is clearly seen in the major axis
Keck/ESI velocity profile shown in the top panel of
Figure~\ref{fig_vpesi}.  The mean line-of-sight velocity profile
reveals a kinematically-distinct core rotating in a direction opposite
to that of the main galaxy body.  In these data, the counter-rotating
region has an average maximum rotation velocity of $v_{\rm rot}^{\rm
core} = 24$\kms, but is asymmetric with respect to the galaxy center.
We show in \S\,\ref{subsec_2d} that this asymmetry is due to a $\sim
15^{\circ}$ misalignment of the ESI slit with respect to the
major-axis of counter-rotation.  The maximum rotation velocity of the
main galaxy body is larger than that of the counter-rotating region.
We do not reach large enough radius to observe a conclusive turnover
in the main galaxy rotation curve and measure a lower limit to the
main body rotation of $v_{\rm rot}^{\rm gal} \ge 45$\kms.  The
velocity dispersion profile in Figure~\ref{fig_vpesi} is symmetric
with respect to the galaxy center, peaking at $\sigma \sim 100$\kms\
at $r = \pm 3.5''$ and declining gently in the central regions to a
central value of 76\kms.  This central drop in the velocity dispersion
profile is observed in other counter-rotating core galaxies
\citep[e.g.,][]{fal04} and is usually interpreted as the central
region being disk-like and kinematically colder than the outer
regions.  However, this is not the only interpretation as galaxies
without counter-rotating cores often show similar drops in the central 
velocity dispersion \citep{mag98,geh03}.

If the main body of NGC~770 were supported by rotational motion then
the expected ratio of the maximum rotation velocity to central
velocity dispersion would be $v_{\rm rot}/\sigma = 0.7$ given a
measured ellipticity of $\epsilon = 0.33$ \citep{bin87}.  The observed
ratio is $v_{\rm rot}/\sigma = 0.47$, but is a lower limit due to our
limited radial coverage.  It is likely that the main galaxy body of
NGC~770 is rotationally supported, consistent with what is seen in
elliptical galaxies of similar luminosity \citep{dav83}.

The bottom two panels of Figure~\ref{fig_vpesi} show the radial profiles
of the Gauss-Hermite moments $h_3$ and $h_4$.  These parameters
measure the deviations of the line-of-sight velocity profile from a
Gaussian profile.  The $h_3$ parameter measures asymmetric
deviations, while $h_4$ measures symmetric deviations.  The $h_3$
profile in Figure~\ref{fig_vpesi} has the same shape as the velocity
profile, but with opposite sign.  The $h_4$ parameter peaks at
negative values (a flatter than Gaussian velocity profile) where the
counter-rotating velocity returns to zero. A plausible interpretation
of the $h_3$ profile is the presence of two kinematically-distinct
components: in the counter-rotating region, the lower velocity stars
of the main galaxy produce an asymmetric tail in the velocity profile
in a direction opposite that of rotation.  However, this sign
difference between the $h_3$ and velocity profiles is also seen in
galaxies without counter-rotating cores \citep[eg.][]{ems04}.  Thus,
although the higher-order moments suggest the presence of two
kinematic components in the inner region of NGC~770, it is hard to
provide an unambiguous interpretation of these data.

\subsection{Two-Dimensional Kinematics}\label{subsec_2d}

\begin{figure}
\epsscale{1.}
\plotone{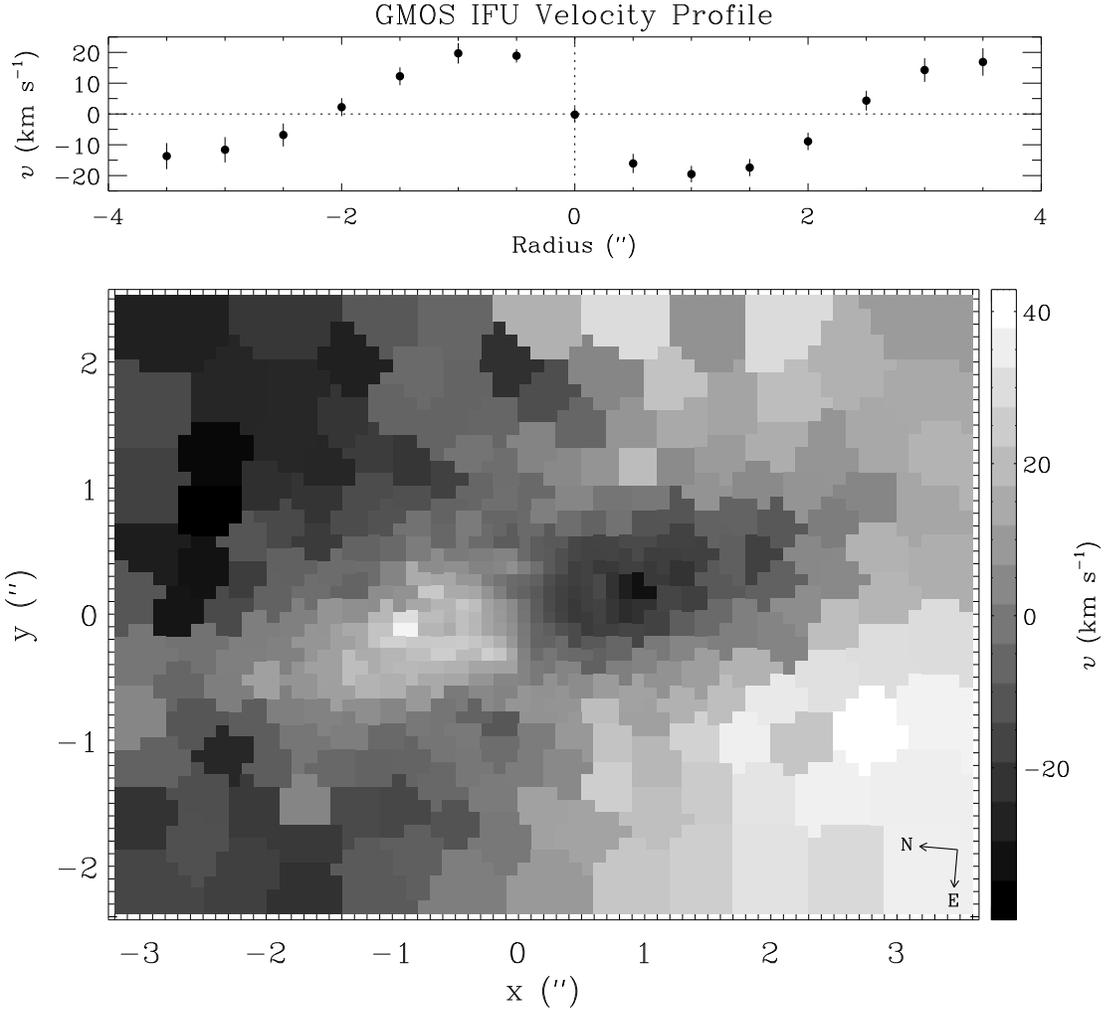}
\caption{({\it Bottom\/})~Stellar velocity ($v$) field for NGC~770
measured with the Gemini GMOS IFU.  The optical center of the galaxy
corresponds to ($x$,~$y$)~=~(0,~0). The orientation of the map is shown in
the lower right corner.  ({\it Top\/})~One-dimensional cut through
the two-dimensional IFU data averaging over a $0.75''$ width slit
centered and aligned along the axis of counter-rotation.  The
major-axis profile is symmetric with respect to the galaxy center; the
maximum rotation velocity of the counter-rotating region is $v_{\rm
rot}^{\rm core} = \pm 21$\kms.
\label{fig_gmos_vel}}
\end{figure}

\begin{figure}
\epsscale{1.}
\plotone{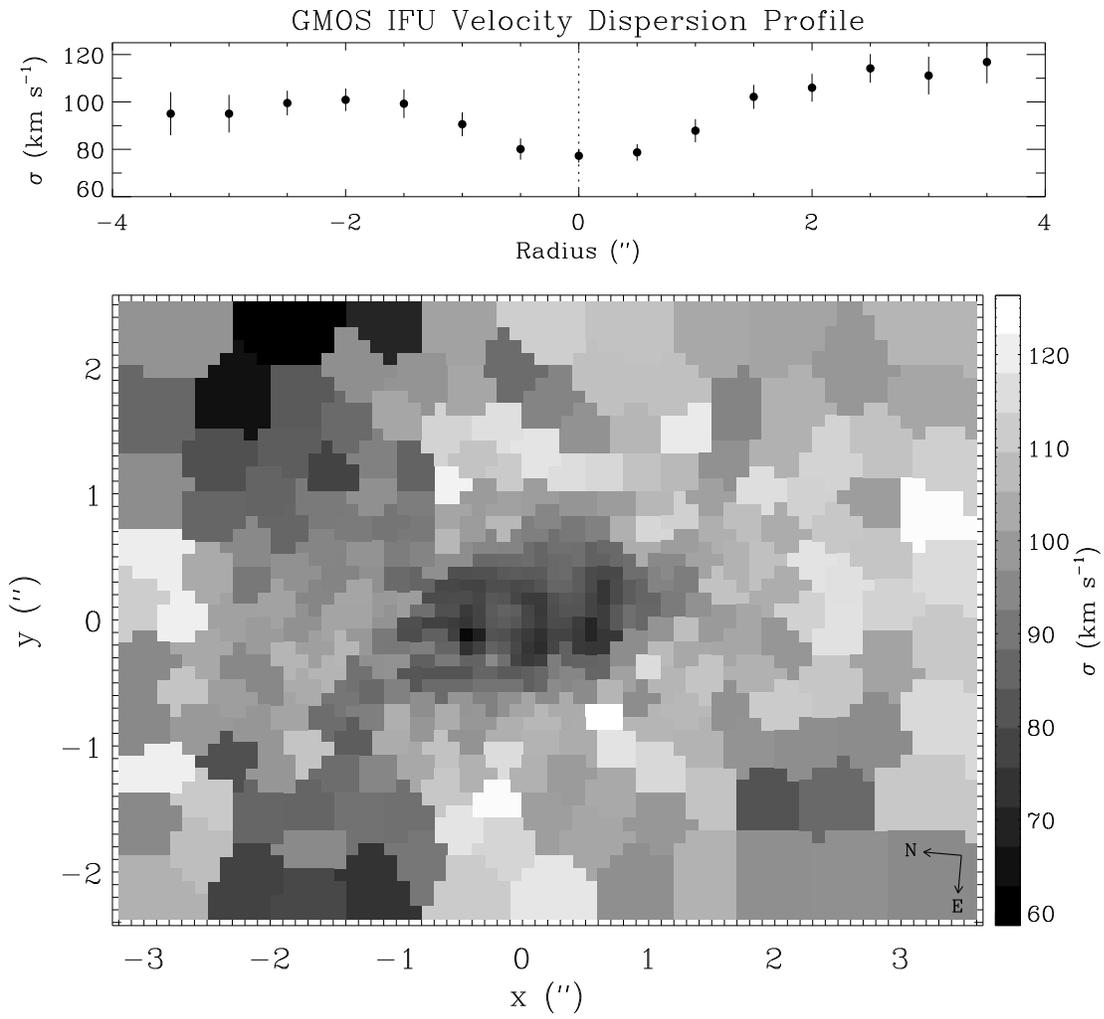}
\caption{Same as Figure~\ref{fig_gmos_vel}, except the
stellar velocity dispersion ($\sigma$) of NGC~770
measured with the Gemini GMOS IFU is shown.
\label{fig_gmos_sig}}
\end{figure}

The two-dimensional Gemini/GMOS IFU velocity
(Figure~\ref{fig_gmos_vel}) and velocity dispersion
(Figure~\ref{fig_gmos_sig}) maps confirm the counter-rotating core in
NGC~770 and provide evidence that stars in this region are moving in
a disk.  The velocity of the counter-rotating core is symmetric with
respect to the photometric center of the galaxy, peaking at a maximum
velocity of $v_{\rm rot}^{\rm core} = \pm 21$\kms\ at a spatial
position of ($x$,~$y$)~=~($0.97''$,~$0.13''$) and ($-0.95''$,~$-0.2''$) in
Figure~\ref{fig_gmos_vel}.  In the top panel of this figure, we show a
one-dimensional velocity profile determined by averaging over a
$0.75''$ wide slit placed along the major axis of counter-rotation.
Unlike the single-slit Keck/ESI data shown in Figure~\ref{fig_vpesi},
the velocity profile is symmetric, peaking at a radial distance of $r=\pm
1''$.  The velocity dispersion map in Figure~\ref{fig_gmos_sig} shows
a colder, lower velocity dispersion central region which is slightly
elongated along the axis of counter-rotation.

We estimate the size of the counter-rotating region by fitting an
ellipse to a zero-velocity contour in Figure~\ref{fig_gmos_vel}
excluding the central half arcsecond.  The best fitting ellipse has a
semi-major axis of $3.5''$ and semi-minor axis of $0.8''$.  The seeing
during these observations was $0.7''$ FWHM. Thus, we are able to put
an upper limit on the scale height of the counter-rotating disk of
$0.8''$ = 130\,pc due to both seeing limitations and the unknown
inclination angle.  This is consistent with the photometric estimates
of the disk size presented in \S\,\ref{sec_disk}.

The position angle of the best-fitting zero-velocity ellipse in
Figure~\ref{fig_gmos_vel} is $15^{\circ}$ relative to the major-axis
of the main galaxy or $20^{\circ}$ East of North.  This misalignment
between the main galaxy body and the counter-rotating core is
consistent with the isophotal twist measured between the inner and
outer galaxy in \S\,\ref{sec_disk}.  This misalignment is sufficient
to explain the asymmetry observed in the single-slit ESI velocity
profile which was placed along the photometric major axis of the main
galaxy body.

\section{Line-Strength Index Profiles}\label{sec_lick}

We compute line-strength indices according to the Lick/IDS system
\citep{wor94} in the wavelength region $4800 - 6000\mbox{\AA}$.  Lick
indices are calculated only for the one-dimensional ESI spectra since the
two-dimensional GMOS IFU data are confined to a narrow spectral
region.  A more detailed description of the procedures to measure
line-strength indices for ESI data is provided in \citet{geh03}.
Briefly, line-strength indices were computed by shifting the ESI
spectra to rest frame wavelengths and convolving with a
wavelength-dependent Gaussian kernel to match the spectral resolution
of the Lick/IDS system.  The size of the kernel ranged between
$8-9\mbox{\AA}$ (40--45 ESI spectral pixels, Gaussian sigma).
Line-strengths were then calculated according to the index definitions
of \citet{wor94}.  Error bars on the indices were computed via
Monte-Carlo simulations which include contributions from photon noise,
read-out noise and pixel-to-pixel correlations due to Gaussian
smoothing.  

\begin{figure}
\epsscale{1.0}
\plotone{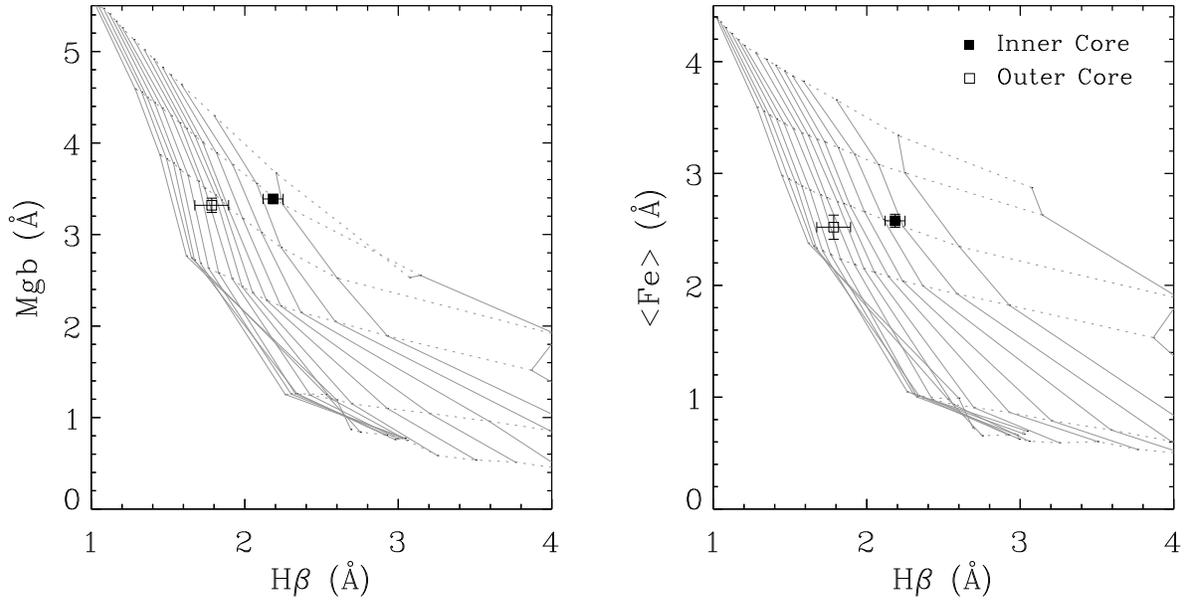}
\caption{Line-strengths Mg\,b and $\langle\rm Fe\rangle$ plotted
against H$\beta$ ({\it left and right, respectively\/}).  The solid
square is the average line-strength measurement of the inner
counter-rotating region ($r\le2''$), the open square is the average
measurement of the outer core ($2'' <r < 4''$).  The solar abundance
models of \citet{tho02} are plotted for ages ranging from 1 to 15~Gyr
in 1~Gyr intervals (solid grey lines) and $[\rm Fe/H] = -2.25$,
$-1.35$, $-0.33$, 0.0, $+0.35$, and $+0.67$\,dex (dotted lines).  In
both panels, lines of constant age are steeper than those of constant
metallicity.
\label{fig_index}}
\end{figure}

We compare the weighted average line-strengths for the inner
($r\le2''$) and outer ($2''<r<4''$) counter-rotating core plotted as
solid and open points, respectively, in Figure~\ref{fig_index}.  We do
not measure line-strengths beyond the counter-rotating region ($r >
4''$) due to insufficient signal-to-noise.  However, because the ESI
slit was slightly misaligned with respect to the counter-rotating
disk, the outer measurement will contain more light from the main galaxy as
compared to the inner measurement.  To determine the
luminosity-weighted stellar ages and metallicities, we plot the Mg\,b
and $\langle \rm Fe \rangle$ (where $\langle \rm Fe \rangle \equiv$
(Fe5270 + Fe5335)/2) indices against H$\beta$ in
Figure~\ref{fig_index}.  We compare our line-strength indices to the
single-burst solar metallicity stellar population models of
\citet*{tho02}.  The best fitting ages and metallicities were
determined by simultaneously minimizing the residuals between the
observed line-strengths and the predicted Mg\,b, $\langle \rm Fe
\rangle$, and H$\beta$ indices from the \citeauthor{tho02}~solar
abundance models.  The best fitting age and metallicity for the inner
counter-rotating region is $3 \pm 0.5$\,Gyr and $[\rm Fe/H] =
0.2\pm0.2$\,dex, and for the outer core region $8\pm 2$\,Gyr and $[\rm
Fe/H] = -0.1\pm 0.2$.  Although the absolute ages and metallicities
are subject to systematic errors in the models, the {\it relative\/}
ages between the two components are more secure.

The difference between the line-strengths measured in the inner and
outer core may be due to a stellar population gradient within the
counter-rotating region or represent contamination from the main
galaxy in the outer measurement.  We favor the latter interpretation
since the age and metallicity for the outer core agree with the global
line-strength measurements presented by \citet{cal03} of NGC~770.  The
outer core is consistent with the average age and metallicity measured
for a sample of dwarf elliptical galaxies of comparable and fainter
luminosities presented by \citet{geh03}.  The age of the inner
counter-rotating region is younger than for typical dwarf elliptical
galaxies.  The stronger H$\beta$ index in the inner core compared to the
outer region suggests that the counter-rotating component contains a
younger stellar population relative to the main galaxy body.

\section{The Environment of NGC 770}\label{sec_environ}

NGC~770 is a low-luminosity elliptical galaxy companion to the large
spiral galaxy NGC~772.  The parent galaxy NGC~772 has an absolute
magnitude $M_B = -21.6$ and is isolated, having no neighbors brighter
than $M_B \le -19$ within 1000 \kms\, and $1.75^{\circ}$ (1\,Mpc) in
the NASA/IPAC Extragalactic Database\footnote{This research has made
use of the NASA/IPAC Extragalactic Database (NED) which is operated by
the Jet Propulsion Laboratory, California Institute of Technology,
under contract with the National Aeronautics and Space
Administration.}.  In addition to NGC~770, \citet{zar97} identified
two other fainter ($M_B \sim -16$) satellites in this system at
projected distances $\sim 400$\,kpc.  Additional fainter satellite
galaxies are likely present in this system.  NGC~770 is the closest
and brightest satellite galaxy to NGC~772 with an absolute magnitude
of $M_B = -18.2$ and a projected distance of $r_p = 3.5'=32$\,kpc.
The physical separation of NCG~770 to its parent galaxy is larger than
that of the Milky Way and the Sagittarius dwarf galaxy
\citep[24\,kpc;][]{iba95}, and is likely comparable to that of
the Large Magellanic Cloud \citep[50\,kpc;][]{fre01} or NGC~205
and M\,31 \citep{vdb00}.

The primary spiral galaxy, NGC~772, is listed in the Atlas of
Peculiar Galaxies \citep[Arp 78;][]{arp66} due to a prominent
asymmetric spiral arm and faint trails of material surrounding both
the primary spiral and the satellite elliptical NGC~770
(Figure~\ref{fig_dss}).  \citet{pig01} showed that the inner 10\,kpc
of this spiral are symmetric, and successfully model the inner spiral
region assuming dynamic equilibrium.  The asymmetric region outside
10\,kpc is due to a single prominent spiral arm which is also the
primary region of on-going star formation; the majority of the
H$\alpha$ flux in this galaxy is emitted in this region \citep{lau89}.

\subsection{The Distribution of HI Gas in the NGC 772/770 System}\label{subsec_hi}

\begin{figure}
\epsscale{1.0}
\plotone{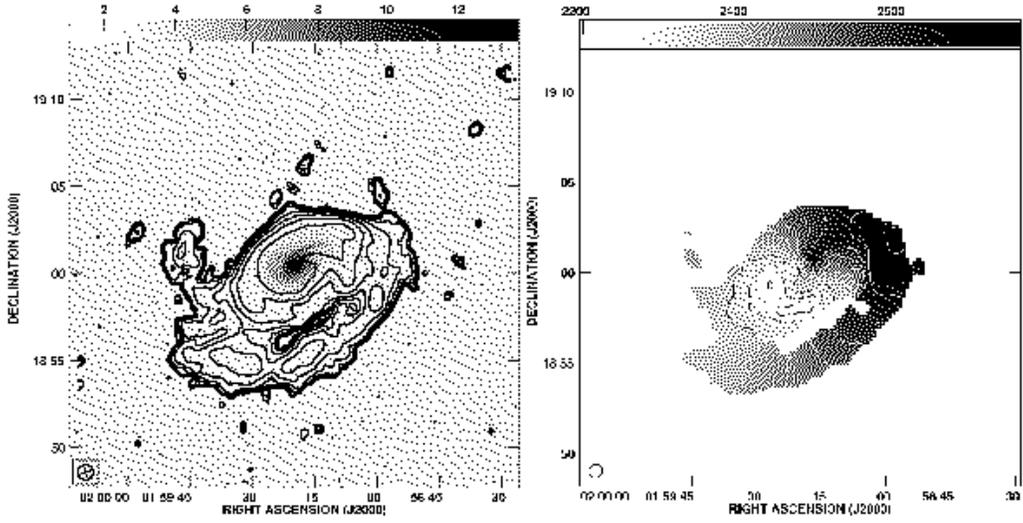}
\caption{({\it Left\/})~Integrated neutral hydrogen HI column density
contours over-plotted on a Digitized Palomar Optical Sky Survey image.
The contours are
between 0.5--$64\times10^{19}$~cm$^{-2}$ incremented by factors of
two.  The beam FWHM of these VLA D-array observations is $\lesssim1'$,
indicated in the lower left corner.  ({\it Right\/})~HI velocity map with
iso-velocity contours plotted every 50\kms\ over the range 2250--2700\kms.
These figures were provided by M.\,Iyer and C.\,Simpson.
\label{fig_hi}}
\end{figure}

The neutral hydrogen (HI) gas in the NGC~772/770 system extends well
beyond the optical radius of NGC~772 and is highly disturbed.  This
system was observed in the 21\,cm line by \citet{iye01} with the Very
Large Array (VLA) in the D-configuration.  The observing set-up and
reduction procedures are similar to those discussed in \citet{iye04}.
The HI flux density map shown in the left panel of Figure~\ref{fig_hi}
was first presented as a minor merger system in the HI Rogues Gallery
\citep{hib01}.  The corresponding HI velocity map in the right panel
of this figure was kindly provided to us by C.\,Simpson.  The total HI
mass in this system is $2.6\times10^{10}\Msun$ \citep{bro97}.

The integrated HI distribution in the NGC~772/770 system is not
centered on the primary galaxy, instead it is skewed toward the
satellite galaxy NGC~770 and extends well beyond the projected
distance of the satellite.  The HI velocity map in the right panel of
Figure~\ref{fig_hi} shows regular rotation in the inner region of the
primary galaxy and a velocity gradient along the outer arm of
material.  A deficiency in the HI distribution is seen in the region
of NGC~770.  Although HI has been detected in NGC~770 \citep{huc89},
the integrated flux density is below the detection threshold of the
VLA observations.  This deficiency can be interpreted as either a hole
in the HI distribution, or the inner region of an HI arm which has
wrapped around the primary galaxy.  The asymmetry of the HI
distribution requires a recent interaction, we discuss possible
scenarios below.

\section{The Influence of Interactions on NGC~770 and 
its Counter-Rotating Core}\label{sec_discuss}

The discovery of a counter-rotating core in NGC~770 is unusual as this
host is not the primary galaxy in this system and it lies in a region
of on-going tidal interactions.  Ongoing star formation in the
asymmetric arm of the primary galaxy NGC~772, and the short dynamical
lifetime of such a structure requires a triggering interaction in the
past few 100\,Myrs \citep{elm91,pig01}.  The NGC~772/770 system is
isolated---the lack of any other massive objects in this system
strongly suggests that this most recent interaction involved NGC~772
and NGC~770.  The presence of a faint optical trail of material close
to satellite galaxy NGC~770 (Figure~\ref{fig_dss}) and its position
relative to the HI gas distribution (Figure~\ref{fig_hi}) support this
hypothesis.  The mass ratio between these two galaxies is 20:1,
assuming they have similar mass-to-light ratios.  Careful modeling is
required to understand the details of this interaction.

Although it is likely that NGC~772 and NGC~770 are undergoing (or have
recently undergone) tidal interaction, the counter-rotating core in
NGC~770 could not have been formed during these most recent
interactions.  The inferred age of the counter-rotating core in the
satellite galaxy NGC~770 is 3\,Gyr, longer than a reasonable estimate
for the orbital period of NGC~770.  The radial velocity difference
between between NGC~772 and NGC~770 is 70\kms.  Assuming a transverse
velocity comparable to the measured radial velocity, the space
velocity of NGC~770 is $\sqrt{3} \times 70 \sim 120$\kms.  If the
projected distance of NGC~770 is the radius of its orbit around
NGC~772, then its orbital period is 1.5\,Gyr.  If NGC~770 is currently
on an unbound orbit, it would have been roughly 500\,kpc from NGC~772
at the time of the counter-rotating core's formation.  Thus, the
counter-rotating core in NGC~770 was formed somewhere between the
outskirts of this group or closer to the primary galaxy NGC~772, two
to three orbits ago.

The lack of other large galaxies in this region limit the possible
formation scenarios for the formation of the counter-rotating core in
NGC~770.  We discuss several possibilities: (1) NGC~770 accreted a
small dwarf galaxy during a very minor merging event, (2) NGC~770
accreted material from the primary galaxy during earlier orbital
interactions, or (3) the counter-rotating region was formed via
angular momentum transfer to the outer envelope of the galaxy via
harassment.  The age difference between the core and the main body of
NGC~770 (\S\,\ref{sec_lick}) suggest that the counter-rotating region
was not formed concurrently with the main galaxy.  The absolute age of
the counter-rotating region of 3\,Gyr suggests that it was not formed
during the interactions which caused the disturbances seen in the
primary spiral galaxy NGC~772 since these features have dynamical
lifetimes of 100\,Myr.  Observational evidence presented in
\S\,\ref{sec_disk} and \S\,\ref{subsec_2d} suggests that the
counter-rotating stars in NGC~770 are moving in a disk.  Thus, any
plausible formation mechanism must explain the counter-rotation, the
age difference between the core and main galaxy body as well as the
presence of a central disk.

A minor merging event which did not destroy NGC~770 but did provide
material with very different angular momentum can explain the observed
counter-rotating core in this galaxy.  Evidence that the
counter-rotating stars rotate in a disk suggests that the merged
galaxy contained a significant gas fraction.  If the core was formed
in such a scenario, the galaxy which merged with NGC~770 would
necessarily be a small dwarf galaxy.  In \S\,\ref{sec_disk}, we
estimate that the inner disky region of NGC~770 contains $\sim3.5$\%
of the total galaxy light.  If we attribute this to the
counter-rotating component, it corresponds to an absolute luminosity
of $M_V \sim -15$ or a mass of $\sim 10^{8}\Msun$ modulo the unknown
gas fraction in such a progenitor and the efficiency of star formation
in low-mass mergers.  

The abundance of HI gas in the NGC~772/770 system makes it possible
that the counter-rotating core formed via gas accretion from the
primary galaxy.  It appears from Figure~\ref{fig_hi} that NGC~770 is
affecting the HI gas distribution of NGC~772.  However, these current
interactions have a dynamical timescale that is much shorter than the
inferred age of the counter-rotating disk.  Thus, this gas accretion
would have had to have occurred two to three orbits ago in NGC~770's
motion around its primary galaxy.  Although it may be possible, it is
unlikely that gas from the primary galaxy would be accreted by the
smaller perturbing galaxy far from the center of mass of the system.

NGC~770 is the faintest known galaxy to host a core which
counter-rotates with respect to the main galaxy kinematics.
\citet{der04} propose an alternative formation mechanism for
kinematically-distinct cores in low luminosity ellipticals in which
angular momentum is transfered to the halo of a dwarf galaxy during
flyby encounters with more massive galaxies.  These authors present
observations of two dwarf elliptical galaxies ($M_B \sim -17$) with
cores rotating in the same direction as the main galaxy body, but with
slower rotation speeds.  In these two systems, the kinematic
decoupling observed between the core and halo can be explained via
galaxy interactions.  In the case of NGC~770, the maximum angular
momentum transfer during its current encounter between NGC~770 and its
parent galaxy is $\sim 2$\kms (assuming a relative velocity of 120
\kms, separation distance of 32\,kpc, and primary galaxy mass
$2\times10^{11}\Msun$; c.f.~Eqn.\,7 of \citet{der04}).  To explain the
counter-rotating core in NGC~770, which rotates with a maximum
velocity of 21\kms\ in a direction opposite the main galaxy, would
require either an unreasonably large number of orbits at the current
separation, or fewer orbits at a closer physical separation.  Thus,
interactions are an unlikely formation mechanism for the
counter-rotating core in NGC~770.

Of the above formation models, we favor the scenario in which the
counter-rotating core in NGC~770 was formed during a minor merging
event with a smaller dwarf galaxy.  \citet{der04} have noted that
dwarf-dwarf merging is unlikely in galaxy-rich groups and clusters due
to large relative velocities between galaxies.  In the low galaxy
density environment of NGC~770, the low relative velocities ($\sim
100$\kms) suggests that NGC~770 could have captured a smaller dwarf
galaxy for impact parameters less than a few tens of kpc.  If the
merging scenario is correct, NGC~770 is an excellent example of
merging at dwarf-sized galaxy scales.

\section{Summary}\label{sec_conc}

We present evidence for a counter-rotating core in the low-luminosity
elliptical galaxy NGC~770 based on Keck/ESI and Gemini/GMOS IFU
spectroscopy.  The combined one- and two-dimensional kinematics argue
for a kinematically-distinct core rotating counter to the main galaxy
body with maximum rotation speed of $21$\kms.  The counter-rotating
region is misaligned with the main body of the galaxy by $15^{\circ}$.
We present evidence that stars in the counter-rotating region are
confined to a disk with radius $4''$ (0.6\,kpc) and we estimate an
upper limit to the scale disk height of $0.8''$ (0.1\,kpc).  We
compute an age and metallicity of the inner counter-rotating region of
$3 \pm 0.5$\,Gyr and [Fe/H] = $0.2\pm 0.2$\,dex, based on Lick
absorption-line indices.  We discuss several possible formation
scenarios for the counter-rotating core in NGC~770 and favor one in
which the core was formed in a minor merging event with a smaller
dwarf galaxy.  If this scenario is correct, it represents one of the
few known examples of merging between two dwarf-sized galaxies.

\acknowledgments

We are grateful to M.~Iyer and C.~Simpson (FIU) for kindly providing
the VLA HI map of NGC~772/770 before publication.  We thank Jeremy
Darling and Francois Schweizer for fruitful discussions regarding this
work.  M.~G.~is supported by NASA through Hubble Fellowship grant
HF-01159.01-A awarded by the Space Telescope Science Institute, which
is operated by the Association of Universities for Research in
Astronomy, Inc., under NASA contract NAS 5-26555.  P.~G.~acknowledges
support from NSF grant AST-0307966.  Based in part on observations
obtained at the Gemini Observatory (Program ID GN-2003B-Q-42), which
is operated by the Association of Universities for Research in
Astronomy, Inc., under a cooperative agreement with the NSF on behalf
of the Gemini partnership: the National Science Foundation (United
States), the Particle Physics and Astronomy Research Council (United
Kingdom), the National Research Council (Canada), CONICYT (Chile), the
Australian Research Council (Australia), CNPq (Brazil) and CONICET
(Argentina).




\clearpage

\begin{deluxetable}{lcccccccccc}
\tabletypesize{\scriptsize}
\tablecaption{NGC 770: Basic Properties}
\tablewidth{0pt}
\tablehead{
\colhead{Name} &
\colhead{$\alpha$ (J2000)} &
\colhead{$\delta$ (J2000)} &
\colhead{Type} &
\colhead{$m_V$} &
\colhead{$A_V$} &
\colhead{$M_{V,0}$}&
\colhead{$\epsilon$}&
\colhead{$r_{\rm eff}$} &
\colhead{$\mu_{V,\rm eff}$}&
\colhead{$n_{\rm serc}$} \\
\colhead{}&
\colhead{(h$\,$:$\,$m$\,$:$\,$s)} &
\colhead{($^\circ\,$:$\,'\,$:$\,''$)} &
\colhead{}&
\colhead{}&
\colhead{}&
\colhead{}&
\colhead{}&
\colhead{[$''$ (kpc)]}&
\colhead{(mag arcs$^{-2}$)}&
\colhead{}
}
\startdata
NGC~770  & 1:59:13.6  & 18:57:17  & E3: & 13.9 &   0.24 & $-18.9$  & 0.33 & 5.2 (0.83) &  19.5 &  2.3 \\ 
\enddata
\tablecomments{Galaxy classification taken from \citet{bin85}; the
``:'' indicates uncertainty in the classification.  The apparent
magnitude is determined from Keck/ESI $V$-band imaging inside an
$r<40''$ aperture.  The absolute magnitude, $M_{V,0}$, assumes a
distance modulus of $(m - M)_0 = 32.6$ and is corrected for foreground
reddening according to \citet{sch98}.}
\end{deluxetable}

\begin{deluxetable}{lcccccccccc}
\tabletypesize{\scriptsize}
\tablecaption{NGC 770: Line-Strength Indices}
\tablewidth{0pt}
\tablehead{

\colhead{} &
\colhead{H$\beta$}&
\colhead{Mgb}&
\colhead{$\langle \rm Fe\rangle$} \\
\colhead{}&
\colhead{($\mbox{\AA}$)}&
\colhead{($\mbox{\AA}$)}&
\colhead{($\mbox{\AA}$)}
}
\startdata
Inner core ($r<2''$)      & $2.18 \pm 0.07$ & $3.39 \pm 0.04$ & $2.58 \pm 0.06$ \\ 
Outer core ($2''<r<4''$)& $1.78 \pm 0.11$ & $3.32 \pm 0.08$ & $2.52 \pm 0.11$ \\
\enddata
\tablecomments{\,Line-strength indices are determined according to the
definitions of \citet{wor94}.  Error bars are
computed via Monte-Carlo simulations as described in
\S\,\ref{sec_lick}.  The combined iron index is defined as $\langle \rm
Fe\rangle \equiv (\rm Fe5270 + \rm Fe5335) / 2$.}
\end{deluxetable}

\clearpage

\newpage
 \end{document}